\documentclass[final,5p,twocolumn,times,sort&compress]{elsarticle}

\def\preprintdate{June 2019}
\def\tfrac#1#2{{\textstyle{{#1}\over {#2}}}}
\def\sheq{\hskip -6pt &=& \hskip -6pt}
\def\sleftrightarrow{\hskip -6pt &\leftrightarrow& \hskip -6pt}

\def\al{\alpha}
\def\be{\beta}
\def\ga{\gamma}

\def\et{\eta}

\def\la{\lambda}

\def\ph{\phi}

\def\ch{\chi}

\def\om{\omega}

\def\De{\Delta}

\def\La{\Lambda}

\def\cA{{\cal A}}
\def\cB{{\cal B}}

\def\cl{{\cal L}}

\def\half{{\textstyle{1\over 2}}}
\def\ol{\overline}
\def\prt{\partial}

\def\lsim{\mathrel{\rlap{\lower4pt\hbox{\hskip1pt$\sim$}}
    \raise1pt\hbox{$<$}}}
\def\gsim{\mathrel{\rlap{\lower4pt\hbox{\hskip1pt$\sim$}}
    \raise1pt\hbox{$>$}}}

\def\etal{{\it et al.}}

\newcommand{\beq}{\begin{equation}}
\newcommand{\eeq}{\end{equation}}
\newcommand{\bea}{\begin{eqnarray}}
\newcommand{\eea}{\end{eqnarray}}
\newcommand{\rf}[1]{(\ref{#1})}
\newcommand{\nn}{\nonumber\\}

\def\po{P^0}
\def\pob{\ol{P^0}}

\def\kk{K^0}
\def\dd{D^0}
\def\bd{B_d^0}
\def\bs{B_s^0}

\def\xid{\xi^{\hskip-1pt(d)}{}}

\def\ha{(\widehat{k}_a){}}
\def\kam{(k_a^{(d)})^{\mu_1\mu_2\ldots\mu_{d-2}}}
\def\kal{(k_a^{(d)})_{\mu_1\mu_2\ldots\mu_{d-2}}}
\def\kaf{(k_a^{(5)}){}}
\def\kfk{(k_{a,K}^{(5)}){}}
\def\kfd{(k_{a,D}^{(5)}){}}
\def\kfb{(k_{a,B_d}^{(5)}){}}
\def\kfbs{(k_{a,B_s}^{(5)}){}}
\def\kfp{(k_{a,P}^{(5)}){}}
\def\ktp{(k_{a,P}^{(3)}){}}

\begin{document}

\begin{frontmatter}

\title{
Searching for CPT Violation with Neutral-Meson Oscillations 
}

\author{Benjamin R.\ Edwards and V.\ Alan Kosteleck\'y}

\address{Physics Department, Indiana University, 
Bloomington, Indiana 47405, USA}

\address{}
\address{\rm 
\preprintdate;
published as Phys.\ Lett.\ B {\bf 795}, 620 (2019)
}

\begin{abstract}

A general technique is presented for treating
CPT violation in neutral-meson oscillations.
The effective field theory for a complex scalar 
with CPT-violating operators of arbitrary mass dimension is incorporated 
in the formalism for the propagation and mixing of neutral mesons.
Observable effects are discussed,
and first measurements of CPT-violating operators of dimension five
are extracted from existing experimental results.

\end{abstract}

\end{frontmatter}

The four neutral mesons
$\kk$,
$\dd$,
$\bd$,
$\bs$
and their antiparticles
play a central role in investigations of fundamental symmetries.
Although electrically neutral,
each of these mesons carries nonzero flavor.
The weak interactions mix each meson with its antiparticle,
thereby inducing flavor oscillations
that can serve as a sensitive interferometer
in searches for physics beyond the Standard Model (SM)
and General Relativity (GR).
The existence of CP violation in nature was uncovered
using this technique 
\cite{ccft64}.

Among the fundamental symmetries
that are accessible to testing with neutral-meson oscillations
is CPT invariance.
This symmetry is guaranteed to hold 
in any relativistic quantum field theory,
including the SM
\cite{cpt}. 
Interest in hypothetical violations of CPT invariance
has increased in recent years,
stimulated by the realization that they may arise 
in an underlying theory unifying quantum physics and gravity
such as strings
\cite{kp91}.
The advent of a comprehensive framework 
for describing CPT violation
at the level of effective field theory,
the Standard-Model Extension (SME)
\cite{kp95,ck97,ak04},
has led to numerous experimental investigations
\cite{ktev,focus,babar,kloe,d0,lhcb}
based on specific and detailed SME predictions
for experimental observables for CPT violation
in neutral-meson systems
\cite{ak98,ak00,ak01,kvk10,tv15,ks16,ar17}.

The propagation and oscillation of a neutral meson
can be described using a 2$\times$2 effective hamiltonian $\La$
acting in the space of meson and antimeson quantum states.
In this formalism, 
CPT violation is controlled by a complex parameter,
denoted here by $\xi$,
that is proportional to the difference between the diagonal elements of $\La$. 
Using this approach,
early analyses studying CPT violation in neutral-meson oscillations
began over 50 years ago
\cite{historical}.
These efforts preceded the establishment of the SM 
and relied primarily on generic arguments at the level of quantum mechanics.
In this context,
it seemed natural to assume that the parameter $\xi$
controlling CPT violation is a single complex number
that is a universal constant quantity for all meson species.
However,
the widespread acceptance of quantum field theory and the advent of the SM 
has radically changed this understanding.
We now know that the SM provides an excellent description of nature
and that small deviations from it can be described
using the tools of effective field theory
\cite{sw}.
For CPT violation,
adopting this methodology yields the SME.
In this comprehensive framework,
it turns out that the parameter $\xi$ 
can depend on the meson flavor and,
crucially,
{\it must} 
depend on the direction and magnitude of the meson 3-momentum 
and on the meson energy
\cite{ak98}.
Moreover,
all laboratories represent noninertial frames
due to the rotation of the Earth,
which implies that experimental observables involving $\xi$ 
also depend on the location of the laboratory on the Earth's surface
and on the sidereal time
\cite{ak98}.
The phenomenological description of CPT violation
is therefore considerably more interesting and involved
than the historical approach would suggest.
Indeed,
experiments that traditionally were regarded
as measuring the same quantity
turn out in fact to be measuring distinct phenomena.

The essential idea underlying the SME
is to extend the SM coupled with GR
to an effective field theory 
that includes all operators violating Lorentz symmetry
\cite{reviews}.
In realistic effective field theory,
any operators violating CPT violate Lorentz symmetry as well
\cite{ck97,owg02},
so the SME also provides a general framework
for studying CPT violation.
Terms in the SME beyond the SM and GR can be organized as a series
of operators of increasing mass dimension $d$, 
each controlled by a coefficient that determines the magnitude of effects 
and that is the target of experimental measurements
\cite{tables}.
The minimal SME is the restriction of the full SME to operators of $d\leq 4$,
which in Minkowski spacetime produces a renormalizable theory.
Nonminimal operators with $d\geq 5$ also play a central role
in formal contexts including studies of 
string theory
\cite{kp91,ks89},
causality and stability
\cite{causality},
and Riemann-Finsler geometry
\cite{ak11,finsler,ek18},
in searches for Lorentz-invariant geometric extensions of GR
including torsion
\cite{torsion}
and nonmetricity
\cite{nonmetricity},
and in phenomenological investigations 
including supersymmetric models 
\cite{susy}
and noncommutative quantum field theories 
\cite{ncqft}.
The key notion in the present context
is that the SME can be used to predict specific observable signals
for CPT violation in the neutral-meson systems.

To date,
all SME analyses of CPT violation in neutral-meson oscillations 
have been performed using the minimal SME.
In this work,
we develop techniques to handle also CPT-violating operators 
of arbitrary nonminimal dimension $d$.
This permits a comprehensive treatment of CPT violation 
in meson propagation and mixing,
which can be used to analyze experimental data
in any of the neutral-meson systems. 
The primary idea is to treat a meson-antimeson system
using CPT-violating scalar effective field theory
\cite{ek18},
which bypasses many of the complications
associated with the nonminimal sector of the SME.
In what follows,
we present the theory 
and then discuss some applications to experiments.

The Schr\"odinger time evolution 
of a linear combination of the wave functions
for a generic neutral meson $\po$ and its antiparticle $\pob$
is governed by a $2\times2$ effective hamiltonian $\La$.
The physical propagating mesons are the eigenstates of $\La$,
which play a role paralleling that of the normal modes 
in a two-dimensional mechanical oscillator
\cite{osc}.
The eigenvalues of $\La$ are complex numbers 
$\la_a \equiv m_a - \half i \ga_a$ and
$\la_b \equiv m_b - \half i \ga_b$,
where $m_a$, $m_b$ are the physical masses 
and $\ga_a$, $\ga_b$ are the corresponding decay rates.
It is convenient to define also the quantities
$\la = \la_a + \la_b$ and $\De \la = \la_a - \la_b$.

Independent of phase conventions
and of the size of CPT violation,
the effective hamiltonian $\La$ can be expressed as
\cite{ak01}
\beq
\La =
\half \De\la
\left(\begin{array}{cc}
U + \xi
&
VW^{-1}
\\
& \\
VW
&
U - \xi
\end{array}\right),
\label{uvwx}
\eeq
where
$U \equiv \la/\De\la$,
$V \equiv \sqrt{1 - \xi^2}$.
The complex parameter $\xi$ controls CPT violation
and is determined by the difference $\De\La$ of diagonal elements of $\La$,
while the modulus $w$ of the complex parameter
$W = w \exp (i\om)$ controls T violation.
The phase $\om$ is unobservable.
Relationships between these quantities 
and various other parametrizations found in the literature
are given in Ref.\ \cite{ak01}.

The explicit form of $\xi$ is determined by the underlying theory,
and it can be found by direct calculation
in the context of effective field theory.
In the minimal SME,
the expression for $\xi$ is known 
at leading order in the coefficients for CPT violation
\cite{ak98}.
However,
incorporating the nonminimal sector of the SME
is more challenging.
In the strong and electromagnetic sectors,
a technique has recently been devised
for constructing all nonminimal terms 
including both propagation and interaction 
\cite{kl19},
but the full SME is yet to appear in the literature.
In the present work we introduce a different approach 
that avoids the complexities of the nonminimal SME,
instead using scalar effective field theory
to determine key features of $\xi$.
The idea is to treat the corrections to the neutral-meson propagator
as arising from the CPT-violating operators of arbitrary $d$
that have recently been constructed 
in scalar effective field theory
\cite{ek18}.
This has the advantage of permitting 
the immediate calculation of the explicit form of $\xi$
while isolating for independent derivation 
\cite{lr19}
the specific connection between the coefficients for CPT violation
in the scalar effective field theory
and the standard SME coefficients.

To implement this idea,
we can follow the general perturbative approach 
previously adopted for the minimal SME
but now applied instead to scalar effective field theory.
In the SM limit, 
the effective hamiltonian $\La$ is CPT invariant
and the meson wave functions can be taken as unperturbed states.
In the minimal SME,
the contribution to $\La$ at leading order in coefficients for CPT violation
can be found by evaluating the expectation values
of the CPT-violating operators in the unperturbed states 
\cite{kp95}.
The dominant effects turn out to arise
from CPT-violating corrections to the valence-quark propagators,
while the CPT-violating effects from the sea are suppressed.
This procedure yields the explicit form of $\xi$
at leading order in minimal-SME coefficients
\cite{ak98}.
An analogous methodology can be applied in the present context
of scalar effective field theory,
with the CPT-violating operators being evaluated 
in the unperturbed meson states.
Note that this approach avoids the complexities
of determining the valence-quark and sea contributions 
in the presence of nonminimal SME operators,
instead treating the meson as a point particle
in scalar effective field theory.
In what follows,
we assume for definiteness that the dominant effects at arbitrary $d$
arise from operators correcting the meson propagation
rather than from meson interactions.
This assumption is known to hold in the minimal SME
\cite{ak98}
and appears reasonable for nonminimal interactions,
which either appear suppressed by other couplings 
or are controlled by independent SME coefficients.
A detailed investigation of contributions from nonminimal interactions
is an interesting open topic for future investigation.

Following this approach,
we model the field operator for the $\po$ and $\pob$ mesons
using a complex scalar $\ph$.
The required modifications to the meson propagation 
can then readily be handled using effective field theory,
even though a complete description of the meson behavior 
is challenging to model using only a single field 
in a hermitian Lagrange density 
because the finite meson lifetime implies nonhermitian contributions 
to the effective hamiltonian $\La$.
Explicitly,
the CPT-violating contributions $\cl_{\rm CPT}$
to the Lagrange density describing the propagation of $\ph$ 
in scalar effective field theory can be written as
\cite{ek18}
\beq
\cl_{\rm CPT}
\supset
- \half i \ph^\dagger\ha^\mu \prt_\mu \ph + {\rm h.c.},
\label{lag}
\eeq
where $\ha^\mu$ is constructed as a series containing 
arbitrary even powers of derivatives $\prt_\al$.
We assume here that the CPT violation is perturbative 
and maintains energy-momentum conservation,
which implies $\ha^\mu$ is independent of spacetime position.
The constant coefficients appearing in the series $\ha^\mu$ 
can then be taken hermitian without loss of generality.
Note that the expression $\cl_{\rm CPT}$ preserves the U(1) symmetry
associated with the meson flavor,
as expected for diagonal contributions to the effective hamiltonian $\La$.
In contrast,
hermitian quadratic contributions to the Lagrange density 
simultaneously violating the flavor U(1) and CPT symmetries
reduce to total-derivative terms leaving unaffected the physics. 
Hermitian quadratic terms violating the U(1) symmetry 
while preserving CPT also exist and can affect oscillations,
but they are irrelevant in determining $\xi$
because they contribute only to off-diagonal components of $\La$.
The terms \rf{lag} are therefore the only ones
of relevance for the calculation of $\xi$,
and taking appropriate expectation values in unperturbed meson states
can be expected to yield hermitian corrections to the effective hamiltonian
and hence real contributions to $\De\La$.

In momentum space with the identification
$i\prt_\mu \leftrightarrow p_\mu$,
we can write
\bea
\ha^\mu \sheq
\sum_{d\geq 3}(k_a^{(d)})^{\mu{\al_1}{\al_2}\ldots{\al_{d-3}}}
p_{\al_1}p_{\al_2}\ldots p_{\al_{d-3}} ,
\label{aop}
\eea
where the sum is over odd values of $d$.
To avoid possible issues with an infinite range for this sum,
we can take the number of Lorentz-violating terms to be arbitrary but finite
\cite{km09}.
The quantities
$(k_a^{(d)})^{\mu\al_1\al_2\ldots\al_{d-3}}$
are the coefficients for CPT violation in the scalar effective field theory.
The label $d$ specifies the mass dimension 
of the corresponding operator in $\cl_{\rm CPT}$,
and the coefficients can be taken as
real spacetime constants with dimension $4-d$.
Each coefficient has $d-2$ spacetime indices and is totally symmetric,
so it is convenient to use a symmetric notation for the indices, 
writing the coefficient as $\kam$.
Field redefinitions permit the traces of a coefficient at fixed $d$
to be absorbed into coefficients at lower $d$
\cite{ek18},
so all coefficients can be taken as traceless.
The number of independent traceless components of $\kam$ at fixed $d$ 
is $(d-1)^2$.
We remark in passing that the properties of $\kam$
imply that the mesons propagate along geodesics of a Riemann-Finsler spacetime
\cite{ak11},
which for the case $d=3$ is a Randers spacetime
\cite{gr41}
and for $d\geq 5$ is a $k$ spacetime
\cite{ek18}.
In this picture,
the flavor conversion from $\po$ to $\pob$
corresponds to a transition 
between two partial geodesics with opposite 4-velocities
in Randers or $k$ space,
matching the usual notion in relativistic quantum mechanics
in which an antiparticle is interpreted as a particle 
moving backwards in time with opposite 3-velocity.

Measuring the coefficients $\kam$
is the goal of experiments searching for CPT violation.
The components of the coefficients are frame dependent,
so reporting a measurement requires specifying a chosen inertial frame.
Due to the rotation of the Earth and its revolution about the Sun,
all laboratories on the surface of the Earth are noninertial.
Instead,
the canonical frame used in the literature to report results
is the Sun-centered frame
\cite{sunframe},
for which 
the time coordinate $T$ has origin at the vernal equinox 2000,
and the spatial coordinates $X^J=(X,Y,Z)$
are defined so that $Z$ parallels the Earth's rotation axis,
$X$ is directed from the Earth to the Sun at $T=0$,
and $Y$ is the orthogonal axis in a right-handed system.
The coefficients of experimental interest in the present context 
are therefore the components of $\kam$
evaluated in the Sun-centered frame.

The general formalism \rf{uvwx} is valid for nonrelativistic mesons,
and the quantities $\xi$ and $w$ governing CPT and T violation
are associated with the meson and defined in a comoving frame.
The frame dependence of the coefficients $\kam$
means that some care is required in deriving
an explicit expression for $\xi$.
Several approaches are possible,
all yielding the same results.
We follow here a stepwise derivation inspired by the method 
introduced in Ref.\ \cite{ak98} in the context of the minimal SME.
Since the Sun-centered frame is the canonical choice
for reporting results on $\kam$,
we first consider a meson at rest in this frame.
Results for a boosted meson and in other frames can then be derived
by judicious use of particle and observer Lorentz transformations.
Also,
although any experimental measurement 
can involve coefficients with multiple values of $d$,
for practical purposes it is often useful and conventional 
to restrict attention to only one value of $d$ at a time
\cite{tables}.
We therefore write
\beq
\xi = \sum_{d=3}^\infty \xid,
\eeq
where the sum is over odd values of $d$,
and we present a derivation of $\xid$ at any fixed $d$.

According to the formalism \rf{uvwx},
the value of $\xid$ is related to the difference
of energies of the $\po$ and $\pob$ mesons. 
The leading-order shift of the energy of a meson
at rest in the Sun-centered frame
is given by the expectation value of the perturbation 
to the nonrelativistic hamiltonian.
The scalar in the effective field theory
has a Klein-Gordon kinetic term,
so the relativistic Lagrange density
contains the combination $\ph^\dagger p^2 \ph +\cl_{\rm CPT}(p)$.
The CPT-violating correction to the unperturbed hamiltonian 
for a meson at rest in the Sun-centered frame
can therefore be identified as $-\cl_{\rm CPT}(m)/2m$,
where $m$ is the $\po$ mass.
This constant shift gives equal and opposite contributions
to the effective energies for $\po$ and $\pob$,
which implies that the explicit form of $\xid$
for mesons at rest in the Sun-centered frame is 
\beq 
\xid = \frac {m^{d-3}} {\De\la} 
(k_a^{(d)})^{TT \cdots T},
\quad
{\rm (rest,~Sun{\rm-}centered~frame)}.
\eeq

To obtain a result valid for mesons moving in the Sun-centered frame,
one must perform a particle boost
while leaving unchanged the coefficients.
Noting that the meson 4-velocity $\be^\mu$ takes the form 
$\be^\mu=(1,0,0,0)$ for mesons at rest,
we can write
$(k_a^{(d)})_{TT \cdots T} 
= \be^{\mu_1}\be^{\mu_2}\cdots\be^{\mu_{d-2}}\kal$.
Performing the particle boost changes the meson 4-velocity
to $\be^\mu = \ga (1, \vec \be )$
but leaves the coefficients $\kam$ unchanged.
This shows that the expression for $\xid$ for moving mesons
in the Sun-centered frame is
\bea 
\xid \sheq \frac {m^{d-3}} {\De\la} 
\be^{\mu_1}\be^{\mu_2}\cdots\be^{\mu_{d-2}}\kal,
\nn
&&
\hskip 50pt
{\rm (boosted,~Sun{\rm-}centered~frame)}.
\label{xibscf}
\eea
Note that this result reveals that a moving meson in the Sun-centered frame
is affected by more components of coefficients for CPT violation
than a meson at rest.
Contributions from coefficients with different $d$
come with different powers of $\be^\mu$ and so are physically distinct.
This also confirms that the coefficients can be taken traceless,
as any trace contribution is proportional 
to at least one power of the Minkowski metric
and hence eliminates two or more factors of $\be^\mu$
via the identity $\et_{\mu\nu}\be^\mu\be^\nu = 1$.

The expression \rf{xibscf} is invariant under changes of observer frame,
which transform both the meson velocities
and the coefficients for CPT violation.
In particular,
this means that the explicit expression for $\xid$
in an inertial frame instantaneously comoving 
with a laboratory on the Earth takes the form
\bea 
\xid \sheq \frac {m^{d-3}} {\De\la} 
\be^{\mu_1}\be^{\mu_2}\cdots\be^{\mu_{d-2}}\kal^{\rm lab},
\nn
&&
\hskip 100pt
{\rm (laboratory~frame)},
\label{xil}
\eea
where $\be^\mu$ is now the meson 4-velocity in the laboratory
and the coefficients $\kal^{\rm lab}$ are measured in the laboratory frame.
The equation relating $\kal^{\rm lab}$ in the laboratory 
to $\kal$ in the Sun-centered frame 
involves an instantaneous observer Lorentz transformation,
so the rotation and revolution of the Earth
imply that the coefficients $\kal^{\rm lab}$ 
depend on the time $T$ and on the location of the laboratory.

In practical applications,
it is convenient to perform a time-space decomposition
of the expression \rf{xil}
and then express the results
in terms of coefficients in the Sun-centered frame.
To achieve this,
first write $\be^\mu = \ga (1, \be \hat\be)$
where $\be$ is the magnitude of the meson three-velocity 
in the laboratory frame,
$\hat\be$ is a unit vector along its spatial direction,
and $\ga= 1/\sqrt{(1-\be^2)}$ is the meson boost factor.
Denoting the time index by $t$
and the spatial ones by $j$ with $j=x,y,z$,
we can then decompose $\xid$ in the laboratory frame as
\beq 
\xid = \frac {m^{d-3}\ga^{d-2}} {\De\la} 
\sum_k {}^{(d-2)}C_k \be^k
\hat\be^{j_1}\cdots\hat\be^{j_k}
(k_a^{(d)})_{t\cdots tj_1\cdots j_k}^{\rm lab},
\eeq
where the sum over $k$ spans $0 \le k \le d-2$
and $^{(d-2)}C_k$ is the usual binomial coefficient.

Next,
note that the laboratory boost 
due to the rotation and revolution of the Earth
is of order $10^{-4}$,
so at zeroth order in this small quantity
the transformation between $\kal^{\rm lab}$ in the laboratory 
and $\kal$ in the Sun-centered frame is a rotation. 
Implementing the rotation yields the desired final result,
\beq 
\xid = \frac {m^{d-3}\ga^{d-2}} {\De\la} 
\sum_k {}^{(d-2)}C_k \be^k
\hat\be^{\prime J_1}\cdots\hat\be^{\prime J_k}
(k_a^{(d)})_{T\cdots TJ_1\cdots J_k},
\label{xifinal}
\eeq
where $\hat\be^{\prime J}$ is a unit vector 
with components in the Sun-centered frame given by 
\bea
\hat\be^\prime{}^X 
&=& (\hat\be^x \cos\ch + \hat\be^z \sin\ch )
\cos{\om_\oplus T_\oplus}
- \hat\be^y \sin{\om_\oplus T_\oplus},
\nn
\hat{\be}^\prime{}^Y
&=&
\hat\be^y \cos{\om_\oplus T_\oplus}
+ ( \hat\be^x \cos\ch + \hat\be^z \sin\ch )
\sin{\om_\oplus T_\oplus}, 
\nn
\hat{\be}^\prime{}^Z
&=&
\hat\be^z \cos\ch - \hat\be^x \sin\ch .
\eea
Here,
$\cos\ch \equiv \hat z \cdot \hat Z$
is the projection of the unit vector in the $z$ direction
onto the unit vector in the $Z$ direction.
This projection can be expressed as a trigonometric combination
of the colatitude and orientation of the laboratory.
The frequency 
$\om_\oplus\simeq 2\pi/(23{\rm ~h} ~56{\rm ~min})$
is the Earth's sidereal rotation frequency.
The time $T_\oplus$ is any convenient local sidereal time
differing from the canonical time $T$ in the Sun-centered frame 
by an appropriate adjustment of the time zero.
For example,
one choice for $T_\oplus$ is associated 
with the standard laboratory frame defined in Ref.\ \cite{sunframe}
and is shifted relative to $T$ by an integer number of sidereal days
and an additional amount that depends on the laboratory longitude
according to Eq.\ (43) of Ref.\ \cite{dk16}.
 
The result \rf{xifinal} 
exhibits several interesting features.
As anticipated,
the parameter $\xid = \xid(\be,\hat\be,T_\oplus,\ch)$ for CPT violation
is found to depend explicitly on the magnitude and direction
of the meson velocity, on the sidereal time,
and on geometric factors associated with the location of the laboratory. 
The oscillatory dependence on sidereal time includes components 
ranging from the zeroth to the $(d-2)$th harmonic
in the Earth's sidereal frequency $\om_\oplus$.
For the special case of $d=3$
only the zeroth and first harmonics appear,
in agreement with known results in the context of the minimal SME
\cite{ak98}.
Expanding the expression \rf{xifinal} 
and comparing to the analogous SME result 
reveals that for $d=3$ the connection between 
the coefficient $(k_a^{(3)})_{\mu}$ 
in the scalar effective field theory
and the SME coefficient combination $\De a_\mu$
is comparatively simple:
$(k_a^{(3)})_{\mu} \approx 2\De a_\mu$.
However,
the corresponding matches for higher $d$
are more involved and lie outside our present scope 
\cite{lr19}.

In the nonminimal sector,
the CPT-violating operator of lowest dimension has $d=5$.
The corresponding coefficient is $(k_a^{(5)})_{\la\mu\nu}$,
and its 20 symmetric components
can conventionally be chosen as
$\kaf_{TTT}$,
$\kaf_{TTX}$,
$\kaf_{TTY}$,
$\kaf_{TTZ}$,
$\kaf_{TXX}$,
$\kaf_{TXY}$,
$\kaf_{TXZ}$,
$\kaf_{TYY}$,
$\kaf_{TYZ}$,
$\kaf_{TZZ}$,
$\kaf_{XXX}$,
$\kaf_{XXY}$,
$\kaf_{XXZ}$,
$\kaf_{XYY}$,
$\kaf_{XYZ}$,
$\kaf_{XZZ}$,
$\kaf_{YYY}$,
$\kaf_{YYZ}$,
$\kaf_{YZZ}$,
and $\kaf_{ZZZ}$.
The requirement of tracelessness implies that 
$(k_a^{(5)})_{\mu\al}{}^{\al}=0$.
This can conveniently be used to eliminate 
the four components $\kaf_{\mu ZZ}$, $\mu = T,X,Y,Z$ via 
\beq
\kaf_{\mu ZZ} \equiv \kaf_{\mu TT} - \kaf_{\mu XX} - \kaf_{\mu YY},
\label{trace}
\eeq
thereby leaving 16 independent observable components.

With the conditions \rf{trace} understood,
the expression \rf{xifinal} for $\xi^{(5)}$ takes the form
\bea 
\xi^{(5)} \sheq 
\frac {m^2\ga^3} {\De\la} 
[(k_a^{(5)})_{TTT}
+ 3 (k_a^{(5)})_{TTJ} \be^{\prime J} 
\nn
&&
\hskip 20pt
+ 3 (k_a^{(5)})_{TJK} \be^{\prime J} \be^{\prime K} 
+ (k_a^{(5)})_{JKL} \be^{\prime J} \be^{\prime K} \be^{\prime L} ]
\nn
\sheq
\frac {m^2\ga^3} {\De\la} 
[\cA_0 
+ \cA_1 \cos\om_\oplus T_\oplus
+ \cB_1 \sin\om_\oplus T_\oplus
\nn
&&
\hskip 20pt
+ \cA_2 \cos 2\om_\oplus T_\oplus
+ \cB_2 \sin 2\om_\oplus T_\oplus
\nn
&&
\hskip 20pt
+ \cA_3 \cos 3\om_\oplus T_\oplus
+ \cB_3 \sin 3\om_\oplus T_\oplus] .
\label{xifive}
\eea
The seven amplitudes 
$\cA_0$, $\cA_1$, $\cB_1$, $\cA_2$, $\cB_2$, $\cA_3$, $\cB_3$
of the harmonic oscillations
are functions of the coefficients $(k_a^{(5)})_{\la\mu\nu}$
for CPT violation in the Sun-centered frame,
the meson velocity $\be^j = (\be^x, \be^y, \be^z)$ in the laboratory frame,
and the geometric factors $c \equiv \cos\ch$ and $s\equiv \sin\ch$.
The zeroth harmonic is independent of sidereal time
and has amplitude $\cA_0$ given by
\bea
\cA_0 \sheq
(k_a^{(d)})_{TTT}
\nn
&&
+3(\be^z c - \be^x s)
\big[
\kaf_{TTZ} + (\be^z c - \be^x s)\kaf_{TZZ}
\big]
\nn
&&
+ \tfrac{3}{2}
\big(
(\be^x c + \be^z s)^2 + (\be^y)^2\big)
[\kaf_{TXX} + \kaf_{TYY}] 
\nn
&&
+ \tfrac{3}{2}(\be^z c - \be^x s)
\big(
(\be^x c + \be^z s)^2 + (\be^y)^2\big)
\nn
&&
\hskip 80pt
\times
[\kaf_{XXZ} + \kaf_{YYZ}]
\nn
&&
+ (\be^z c - \be^x s)^3\kaf_{ZZZ} .
\label{amp1}
\eea
The amplitudes $\cA_1$ and $\cB_1$ for the first harmonics are
\bea
\cA_1 \sheq 
3(\be^x c + \be^z s)\kaf_{TTX}  + 3\be^y\kaf_{TTY} 
\nn
&&
+ 6(\be^z c - \be^x s)
\big[
(\be^x c + \be^z s)\kaf_{TXZ}
+ \be^y\kaf_{TYZ}
\big] 
\nn
&&
+\tfrac{3}{4}((\be^x c + \be^z s)^2 + (\be^y)^2)
\nn
&&
\hskip 40pt 
\times
\big[
(\be^x c + \be^z s)
[\kaf_{XXX} + \kaf_{XYY}]
\nn
&&
\hskip 50pt 
+\be^y[\kaf_{XXY} + \kaf_{YYY}]
\big]
\nn
&&
+3(\be^z c - \be^x s)^2
\big[
(\be^x c + \be^z s)\kaf_{XZZ}
+ \be^y\kaf_{YZZ}
\big] 
\nn
\label{amp2}
\eea
and
\bea
\cB_1 \sheq 
3(\be^x c + \be^z s)\kaf_{TTY} - 3\be^y\kaf_{TTX}
\nn
&&
+ 6(\be^z c - \be^x s)
\big[
(\be^x c + \be^z s)\kaf_{TYZ}
- \be^y\kaf_{TXZ}
\big] 
\nn
&&
+\tfrac{3}{4}
\big(
(\be^x c + \be^z s)^2 + (\be^y)^2\big)
\nn
&&
\hskip 40pt 
\times
\big[
(\be^x c + \be^z s)
[\kaf_{XXY} + \kaf_{YYY}]
\nn
&&
\hskip 50pt 
- \be^y [\kaf_{XXX} + \kaf_{XYY}]
\big] 
\nn
&&
+3(\be^z c - \be^x s)^2
\big[
(\be^x c + \be^z s)\kaf_{YZZ}
- \be^y\kaf_{XZZ}
\big].
\nn
\label{amp3}
\eea
The amplitudes $\cA_2$ and $\cB_2$ for the second harmonics are
\bea
\cA_2 \sheq 
\tfrac{3}{2}
\big(
(\be^x c + \be^z s)^2 - (\be^y)^2\big)
[\kaf_{TXX} - \kaf_{TYY}]
\nn
&&
+ 6\be^y(\be^x c + \be^z s)
\big[
\kaf_{TXY} + (\be^z c - \be^x s)\kaf_{XYZ}
\big] 
\nn
&&
+\tfrac{3}{2}(\be^z c - \be^x s)
\big(
(\be^x c + \be^z s)^2 - (\be^y)^2\big)
\nn
&&
\hskip 40pt 
\times
[\kaf_{XXZ} - \kaf_{YYZ}]
\label{amp4}
\eea
and
\bea
\cB_2 \sheq 
-3\be^y(\be^x c + \be^z s)
[\kaf_{TXX} - \kaf_{TYY}]
\nn
&&
+ 3\big(
(\be^x c + \be^z s)^2 - (\be^y)^2\big)
\nn
&&
\hskip 40pt 
\times
\big[
\kaf_{TXY} + (\be^z c - \be^x s)\kaf_{XYZ}
\big] 
\nn
&&
-3\be^y(\be^x c + \be^z s)(\be^z c - \be^x s)
[\kaf_{XXZ} - \kaf_{YYZ}].
\nn
\label{amp5}
\eea
Finally,
the amplitudes $\cA_3$ and $\cB_3$ for the third harmonics are
\bea
\cA_3 \sheq 
\tfrac{1}{4}(\be^x c + \be^z s)
\big(
(\be^x c + \be^z s)^2 - 3(\be^y)^2\big)
\nn
&&
\hskip 40pt 
\times
[\kaf_{XXX} - 3\kaf_{XYY}] 
\nn
&&
+\tfrac{1}{4}\be^y
\big(
3(\be^x c + \be^z s)^2 - (\be^y)^2\big)
\nn
&&
\hskip 40pt 
\times
[3\kaf_{XXY} - \kaf_{YYY}]
\label{amp6}
\eea
and 
\bea
\cB_3 \sheq 
\tfrac{1}{4}\be^y
\big(
3(\be^x c + \be^z s)^2 - (\be^y)^2\big)
\nn
&&
\hskip 40pt 
\times
[3\kaf_{XYY} - \kaf_{XXX}]
\nn
&&
+\tfrac{1}{4}(\be^x c + \be^z s)
\big(
(\be^x c + \be^z s)^2 - 3(\be^y)^2\big)
\nn
&&
\hskip 40pt 
\times
[3\kaf_{XXY} - \kaf_{YYY}].
\label{amp7}
\eea

If sufficient data are taken in a given experiment,
then the expressions \rf{amp1}-\rf{amp7}
reveal that binning in sidereal time permits
the measurement of seven independent linear combinations 
of the 16 observable components of $\kaf_{\la\mu\nu}$.
Since the specific linear combinations depend on the meson boost,
an experiment with a sufficiently broad meson spectrum 
can obtain more independent measurements by binning in momentum as well.
The dependence on geometric factors implies
that distinct experiments with the same meson spectra
can also have different sensitivities.
For any specific meson species,
all 16 components of $\kaf_{\la\mu\nu}$ appear in the above amplitudes
with distinct multiplicative factors,
so each component is therefore independently measurable in principle.
However,
if this separation is infeasible in a given experiment,
then insight about the comparative sensitivities
achieved for different components can nonetheless be obtained
following standard practice in the field
\cite{tables},
by placing constraints on each independent component
taken one at a time with all others set to zero.

Different experiments may prepare mesons in distinct ways.
Some use uncorrelated mesons from various production processes,
while others use correlated ones obtained 
from decays of quarkonia at rest or boosted.
Measurements of coefficients with $d=3$
have been performed with uncorrelated mesons using the 
KTeV
\cite{ktev},
D0
\cite{d0},
FOCUS
\cite{focus},
and
LHCb
\cite{lhcb}
detectors,
while ones with unboosted correlated $\kk$ have been completed at
KLOE
\cite{kloe}
and ones with
boosted correlated $\bd$ mesons at 
BaBar
\cite{babar}.
Other experiments could also achieve interesting sensitivities
to coefficients for CPT violation.
For example,
the Belle II experiment 
\cite{belle}
also involves correlated mesons 
and in principle could obtain competitive constraints 
\cite{ar17}.
Theoretical asymmetries that isolate CPT violation
are discussed in Refs.\ \cite{ak98,ak00,ak01,kvk10,tv15,ks16,ar17},
and investigations of these observables
and other techniques have been adopted 
in the various experimental analyses.
Analogous methodologies can be applied for other values of $d$.
In particular,
since experiments have already studied the zeroth and first harmonics
of the sidereal time for all meson species,
the results reported for coefficients with $d=3$
can be used to deduce constraints for coefficients with higher $d$. 
Note that the general result \rf{xifinal} reveals that
a factor of the boost $\be$ accompanies each appearance
of a spatial index $J$ in any component of $\kal$,
whereas components of the coefficients with $d=3$ 
are limited to at most a single such factor.
Inferring new results in this way thus requires some care.

These ideas can be illustrated using the case of $d=5$,
for which the zeroth and first harmonics
are controlled by the amplitudes \rf{amp1}-\rf{amp3}.
While many components of $\kaf_{\la\mu\nu}$ appear in these amplitudes,
only the components $\kaf_{TTT}$ and $\kaf_{TTJ}$
are accompanied with zero or one power of the boost.
We can therefore convert published measurements 
of coefficients with $d=3$  
into measurements of $\kaf_{TTT}$ and $\kaf_{TTJ}$,
provided we assume that all other observable coefficients
entering the amplitudes \rf{amp1}-\rf{amp3} vanish.
At a cruder level with the boost factors disregarded,
constraints could in principle be extracted also
on the component combinations 
$\kaf_{TYZ}$, 
$\kaf_{TXX}+\kaf_{TYY}$, 
and $\kaf_{XXZ}+\kaf_{YYZ}$
via the amplitude $\cA_0$,
and on the component combinations
$\kaf_{TXZ}+\kaf_{TYZ}$, 
$\kaf_{XXX}+\kaf_{XYY}$, 
and $\kaf_{XXY}+\kaf_{YYY}$
via the amplitudes $\cA_1$ and $\cB_1$.
A complete coverage of all $TJK$ and $JKL$ components of $\kaf_{\la\mu\nu}$
would require a study of the second and third sidereal harmonics
using the amplitudes \rf{amp4}-\rf{amp7},
which to date is lacking in the literature
for every meson species.
Here,
we focus on extracting constraints on 
the components $\kaf_{TTT}$ and $\kaf_{TTJ}$,
leaving other prospective investigations open for future research.

In the theoretical analysis above,
we considered CPT violation in a generic neutral-meson system
using notation that makes no distinction between the different species. 
However,
the behavior of each meson species $\po = \kk$, $\dd$, $\bd$, $\bs$
can in principle be governed by distinct coefficients
$(k_{a,P}^{(d)})_{\mu_1\mu_2\ldots\mu_{d-2}}$
controlling CPT violation.
For instance,
in the case $d=5$ with coefficients $\kfp_{\la\mu\nu}$, 
experiments can report bounds 
on a total of 64 independent observables for CPT violation,
corresponding to the independent traceless components of the coefficients
$\kfk_{\la\mu\nu}$,
$\kfd_{\la\mu\nu}$,
$\kfb_{\la\mu\nu}$,
and $\kfbs_{\la\mu\nu}$
that are required to characterize $d=5$ CPT violation completely
across the four neutral-meson systems.
In particular,
in extracting results for coefficients with $d=5$
from existing experimental results on coefficients with $d=3$,
we can in principle access the 16 components 
$\kfp_{TTT}$ and $\kfp_{TTJ}$
among the 64 independent observable components of $\kfp_{\la\mu\nu}$.

\renewcommand\arraystretch{1.4}
\begin{table}
\caption{
\label{table}
Constraints on $\kfp_{TT\mu}$
for $\kk$, $\dd$, $\bd$, and $\bs$ mesons.
}
\setlength{\tabcolsep}{3pt}
\begin{tabular}{llc}
\hline
\hline
Coefficient & Constraint (GeV$^{-1}$) & Ref. \\
\hline
$	|\kfk_{TTT} - 0.97\kfk_{TTZ}|	$&$	< 2 \times 10^{-24}	$&	\cite{ak98}	\\	
$	|\kfk_{TTX}|,~|\kfk_{TTY}|	$&$	< 1.8 \times 10^{-25}	$&	\cite{ktev}	\\	
$	\kfk_{TTT}	$&$	(-2.4 \pm 4.4) \times 10^{-17}	$&	\cite{kloe}	\\	
$	\kfk_{TTX}	$&$	(1.2 \pm 2.8) \times 10^{-18}	$&	\cite{kloe}	\\	
$	\kfk_{TTY}	$&$	(2.7 \pm 2.7) \times 10^{-18}	$&	\cite{kloe}	\\	
$	\kfk_{TTZ}	$&$	(4.2 \pm 3.0) \times 10^{-18}	$&	\cite{kloe}	\\	[6pt]
$	N^D[\kfd_{TTT} + 0.97\kfd_{TTZ}]	$&$	(-5.1  {\rm ~to~}  8.8) \times 10^{-20} 	$&	\cite{focus}	\\	
$	N^D \kfd_{TTX}, ~N^D \kfd_{TTY} 	$&$	(-6.5  {\rm ~to~} 3.5) \times 10^{-20} 	$&	\cite{focus}	\\	[6pt]
$	N^{B_d}[\kfb_{TTT} - 0.7\kfb_{TTZ}]	$&$	(-1.3 \pm 1.0) \times 10^{-16}	$&	\cite{babar}	\\	
$	N^{B_d}\kfb_{TTX}	$&$	(-3.6 \pm 1.1) \times 10^{-16}	$&	\cite{babar}	\\	
$	N^{B_d}\kfb_{TTY}	$&$	(-4.5 {\rm ~to~}  -0.7) \times 10^{-16}	$&	\cite{babar}	\\	
$	\kfb_{TTT} - 0.8\kfb_{TTZ}	$&$	(-1.3 \pm 17) \times 10^{-20}	$&	\cite{lhcb}	\\	
$	\kfb_{TTX}	$&$	(1.0 \pm 0.8) \times 10^{-19}	$&	\cite{lhcb}	\\	
$	\kfb_{TTY}	$&$	(2.3 \pm 8.1) \times 10^{-20}	$&	\cite{lhcb}	\\	
$	\kfb_{TTT} - 0.7\kfb_{TTZ}	$&$	(-8.0 \pm 5.1) \times 10^{-16}	$&	\cite{ks16}	\\	[6pt]
$	\kfbs_{TTT}	$&$	(1.0 \pm 1.1) \times 10^{-14}	$&	\cite{kvk10}	\\	
$	|\kfbs_{TTX}|,~|\kfbs_{TTY}|	$&$	< 1.1 \times 10^{-15}	$&	\cite{d0}	\\	
$	\kfbs_{TTT} - 0.4\kfbs_{TTZ}	$&$	(-1.6 {\rm ~to~}  8.3) \times 10^{-16}	$&	\cite{d0}	\\	
$	\kfbs_{TTT} - 0.8\kfbs_{TTZ}	$&$	(-1.1 \pm 2.1) \times 10^{-18}	$&	\cite{lhcb}	\\	
$	\kfbs_{TTX}	$&$	(0.5 \pm 1.4) \times 10^{-18}	$&	\cite{lhcb}	\\	
$	\kfbs_{TTY}	$&$	(-1.9 \pm 1.4) \times 10^{-18}	$&	\cite{lhcb}	\\	[2pt]
\hline
\hline
\end{tabular}
\end{table}

The equations implementing the conversion 
from the published $d=3$ results to $d=5$ ones 
can be obtained by comparing 
the amplitudes for $\xi^{(3)}$ obtained from the expression \rf{xifinal}
with the amplitudes \rf{amp1}-\rf{amp3} for $\xi^{(5)}$.
The explicit form of the match is 
\bea
\ktp_T \sleftrightarrow
m^2\ga^2 [1 + 3(\be^z C_\ch - \be^x S_\ch)^2] \kfp_{TTT},
\nn
\ktp_X \sleftrightarrow
3m^2\ga^2 [1 + (\be^z C_\ch - \be^x S_\ch)^2] \kfp_{TTX},
\nn
\ktp_Y \sleftrightarrow
3m^2\ga^2 [1 + (\be^z C_\ch - \be^x S_\ch)^2] \kfp_{TTY},
\nn
\ktp_Z \sleftrightarrow
3m^2\ga^2 [1 + \tfrac{1}{3}(\be^z C_\ch - \be^x S_\ch)^2] \kfp_{TTZ}.
\eea
Some results for $\kfp_{TTT}$ and $\kfp_{TTJ}$
extracted from the existing literature using this match 
are displayed in Table \ref{table}.
The table has four parts,
one for each of the four mesons $\kk$, $\dd$, $\bd$, and $\bs$.
Within each part,
the rows are organized according to the chronological appearance 
of the original constraint for the $d=3$ case.
The first column contains the combinations of components
of $\kfp_{\la\mu\nu}$ for which results can be deduced.
The quantities $N^P$ appearing in some of the combinations
are defined as $N^P = \De m/\De \ga$,
with $\De m = |m_b - m_a|$, $\De \ga = |\ga_a - \ga_b|$
evaluated for the $\po$-meson system in question.
The second column lists the results 
expressed in units of inverse GeV.
In some cases the result is a bound on a magnitude,
in others it is a measurement with standard deviation,
and in yet others it is a range of allowed values,
in accordance with the presentation of the original measurements
for $d=3$ in the literature.
In obtaining these results,
we adopted mean values of the boost factors 
from the published meson spectra.
This procedure could in principle be refined 
by the corresponding experimental collaborations
via detailed reanalysis of the original data 
incorporating the full meson spectra.
The final column provides citations to the source literature 
reporting the original measurements of coefficients with $d=3$.

The results in Table \ref{table} represent 
the first reported sensitivities 
to nonminimal coefficients for CPT violation 
in neutral-meson oscillations.
They extend and complement
sensitivities to $d=5$ spin-independent CPT violation
obtained from analyses of experiments with other systems,
including
charged leptons, protons, neutrons, and neutrinos
\cite{other}.
No comparable effects for photons or gravity
are possible in the context of effective field theory,
where the $d=5$ CPT-violating operators are necessarily spin dependent.
However,
potential sensitivities to quark-sector SME coefficients
controlling $d=5$ spin-independent CPT violation
have been proposed for processes 
such as deep inelastic and Drell-Yan scattering
\cite{kl19,disdy}
and are expected to modify top-quark production and decay
in analogy to known effects in the minimal SME 
\cite{bkl16}.
Spin-independent quark-sector coefficients with $d=5$ can also induce 
phenomenologically viable baryogenesis in thermal equilibrium
\cite{bckp97},
potentially bypassing the necessity for CP violation beyond the SM
and the Sakharov requirement of nonequilibrium processes 
\cite{as67}.
The results obtained in the present work are competitive,
thereby reconfirming the exquisite sensitivity of meson interferometry
to CPT violation.
The prospects are excellent for further advances and potential discovery 
in searches for CPT violation using the neutral-meson systems.

\section*{Acknowledgments}

This work was supported in part
by the United States Department of Energy 
under grant number {DE}-SC0010120
and by the Indiana University Center for Spacetime Symmetries.

\end{document}